# Ultrashort pulsed laser induced complex surface structures generated by tailoring the melt hydrodynamics

F. Fraggelakis [1*], G. D. Tsibidis [1,2*] and E. Stratakis [1,2*]

[1] Institute of Electronic Structure and Laser (IESL), Foundation for Research and Technology (FORTH), N. Plastira 100, Vassilika Vouton, 70013, Heraklion, Crete, Greece

[2] Department of Physics, University of Crete, 71003 Heraklion, Greece

**Correspondence:** fraggelakis@iesl.forth.gr, tsibidis@iesl.forth.gr, stratak@iesl.forth.gr

**Abstract:** We present a novel approach for tailoring the laser induced surface topography upon femtosecond (fs) pulsed laser irradiation. The method employs spatially controlled double fs laser pulses to actively regulate the hydrodynamic microfluidic motion of the melted layer that gives rise to the structures formation. The pulse train used, in particular, consists of a previously unexplored spatiotemporal intensity combination including one pulse with Gaussian and another with periodically modulated intensity distribution created by Direct Laser Interference Patterning (DLIP). The interpulse delay is appropriately chosen to reveal the contribution of the microfluidic melt flow, while it is found that the sequence of the Gaussian and DLIP pulses remarkably influences the surface profile attained. Results also demonstrate that both the spatial intensity of the double pulse and the effective number of pulses per irradiation spot can further be modulated to control the formation of complex surface morphologies. The underlying physical processes behind the complex patterns' generation were interpreted in terms of a multiscale model combining electron excitation with melt hydrodynamics. We believe that this work can constitute a significant step forward towards producing laser induced surface structures on demand by tailoring the melt microfluidic phenomena.

## 1 Introduction

During the past decades, materials' micro and nano fabrication using ultrashort laser pulses have emerged as a key technology, which has contributed to major advances in science, technology and industry [1-6]. Compared to expensive and time consuming techniques, such as photo- or electron beam-lithography [7, 8] that do not provide the necessary specificity and scalability for industrial applications, laser-based processing constitutes an efficient and precise technology to fabricate novel functional materials and devices [9, 10].

Laser surface processing is a rapid and scalable chemical-free technique for surface functionalization that can be applied on almost any kind of material. Laser Induced Periodic Surface Structures (LIPSS) is the most common type of morphology that can be produced via irradiation of solids' surfaces by Gaussian beams [1, 11]. Depending on the laser wavelength, $\lambda_L$, used to fabricate LIPSS, their period $\Lambda$ varies from (i) deep/shallow subwavelength ($\Lambda <<\lambda_L/2$), termed as High Spatial Frequency LIPSS (HSFL) [11, 12] to (ii) subwavelength ($\lambda_L/2< \Lambda< \sim\lambda_L$), coined as Low Spatial Frequency LIPSS (LSFL) [13, 14], to (iii) suprawavelength ($\Lambda>\lambda_L$), called grooves [15-18] or spikes [17, 19], as well as other complex structures [20, 21]. The formation mechanisms of both the LSFL and HSFL structures is attributed to a spatially modulated deposition of the laser energy as a result of the interference of the incident beam and electromagnetic waves scattered on a rough surface in the far- and near-field respectively [1, 14, 22, 23].

Apart from LIPSS formation, an alternative laser processing method, namely the Direct Laser Interference Patterning (DLIP), has recently been introduced to fabricate periodic structures on solid surfaces. DLIP is based on the use of two or more coherent laser beams that interfere and the resulting laser intensity profile irradiates the material to fabricate periodic structures on large areas [24-27]. The imprinted pattern is determined by the angle, the number and the coherence of the interfering beams [28, 29]. The capability to tailor the features of a surface topography by controlling the angle of incidence of the coherent beams offers the capability to form a broad range of surface structures for potential applications [24-27]. Nonetheless, despite the ongoing research and investigation of the various surface patterns, a quantitative analysis of the underlying physical mechanisms behind the DLIP structures' formation is still missing. In some recent works, the role of the induced thermal effects has been investigated [30, 31]. Recently, we have introduced a multiscale theoretical model taking into account the impact of the microfluidic factor as well to interpret the DLIP pattern formation upon double pulse irradiation (DPI) [32]. Theoretical simulations of the periodic patterns formed by two- and four-beam DLIP double pulses, confirmed by experiments, demonstrate the significant role of melt hydrodynamics in the surface structuring process [32].

Despite the significant progress to date, the desired level of control of laser induced structures' characteristics including symmetry, size and hierarchical features formation is far from being accomplished. LSFL and HSFL with multiple axis of symmetry have been only reported as a result of polarisation modification or DPI [19, 21]. Matching the extraordinary variety of structures and functionalities found in nature requires an in-depth understanding of the formation mechanism along with developing the suitable irradiation techniques.

Given the impact of the intensity profile and polarisation in the produced structures, it would be crucial to explore whether appropriate control of the spatiotemporal energy deposition profile would lead to more complex surface morphologies with enhanced application-based functionalities. To the best of our knowledge, no previous study has been conducted on the investigation of the type of topographies that can be produced if double pulses of different spatial intensity are used. In principle, double pulses of Gaussian [33, 34] or DLIP [32] shapes have been previously reported, however the combined effect of Gaussian and DLIP profiles should be important to explore. As pointed out above, the Gaussian beam leads to the formation of LSFL structures, while DLIP yields morphologies that depend on the angle between the coherent beams and the polarisation. Therefore, a thorough investigation of the combined effect of such pulses and inter-pulse delay on surface structures' formation would be important to study. Of further importance is the investigation of the resultant microfluidic movement derived from the phase transition caused under laser irradiation. The application of delayed pulses is crucial to reveal the impact of melt flow, considering that sufficient time is required for hydrodynamical effects to dominate, prior to the action of the second pulse that arrives at the materials' surface and influences the melt flow. For this purpose, the pulse separation here is kept on the order of ~500 ps, which has been proven to be suitable to highlight the role of the melt microfluidic movement [35].

## 2 Experimental and simulation protocols

### a. Laser processing

In the experiments, we utilize a DPI irradiation scheme employing two temporally delayed fs pulses with different intensity distribution. Namely, a pulse with Gaussian distribution (Figure 1-III, indicated as *G*) is combined with a pulse with a 1D or 2D intensity distribution produced by DLIP ( Figure 1-III, indicated as *V*, *H* or *D*). For the experiments, fs pulses, emitted by a Pharos (Light Conversion) laser source with duration of $\tau_p \cong 170$ fs at a wavelength $\lambda_L = 1026$ nm have been used. As shown in Figure 1 a delay line is built to generate double pulses. In particular, the primary pulse is initially divided into two pulses with perpendicular polarisation orientation by a polarizing beam splitter. The energy distribution between the pulses is regulated by a half waveplate (HWP) placed before the polarizing beam splitter (PB). The two generated pulses pass through the arms A and B corresponding to optical paths that differ by 150 ± 1 mm acquiring an interpulse delay of $\Delta\tau$ = 500 ± 3 ps. The two pulses are guided to a computer controlled programmable Spatial Light Modulator (SLM) and via a suitable optical system, comprising two lenses, and interfere onto the sample. The two focusing lenses $f_1$ = 400 mm and $f_2$ = 30 mm are placed on the appropriate distances recombine the beams on the sample giving a spot diameter of 190 μm radius (2w). The working principle of the SLM screen is described in Figure 1 (I). Although the *s*-polarized pulse is reflected from the SLM and maintains its Gaussian shape (Figure 1, *G*), the *p*-polarized one interacts with the SLM screen and is divided into two or four beams acquiring on the sample a profile comprising vertical (Figure 1, *V*), or horizontal (Figure 1, *H*) lines, or a periodic array of dots ( Figure 1, *D*). The periods of DLIP patterns used, $\Lambda_{DLIP}$, were comparable to the laser wavelength. Owing to a rotating HWP placed before the DLIP part, the initial order of the *s*- and *p*- pulse can be inverted enabling one to control whether the Gaussian or the DLIP pulse reaches the sample first. The total energy, $E_{tot}$, for the pair of pulses, the total number of irradiation shots, *NP*, as well as the order of the pulses was varied, as indicated in the experimental section. The energy distribution was 2/3 of $E_{tot}$ for the Gaussian and 1/3 $E_{tot}$ for the DLIP pulse respectively, tuned to maintain a similar effective fluence in the irradiated areas. *NP* was varied in the range of 10 and 500 pulses. Images of the processed surfaces were acquired via Scanning Electron Microscopy (SEM) microscopy, while a Fast Fourier Transformation (FFT) was performed to calculate the periodicities of the induced structures, through the use of the open source software Gwyddion. For all the experiments, mirror polished commercially available 316-stainless steel (provided by RS) substrates have been used.

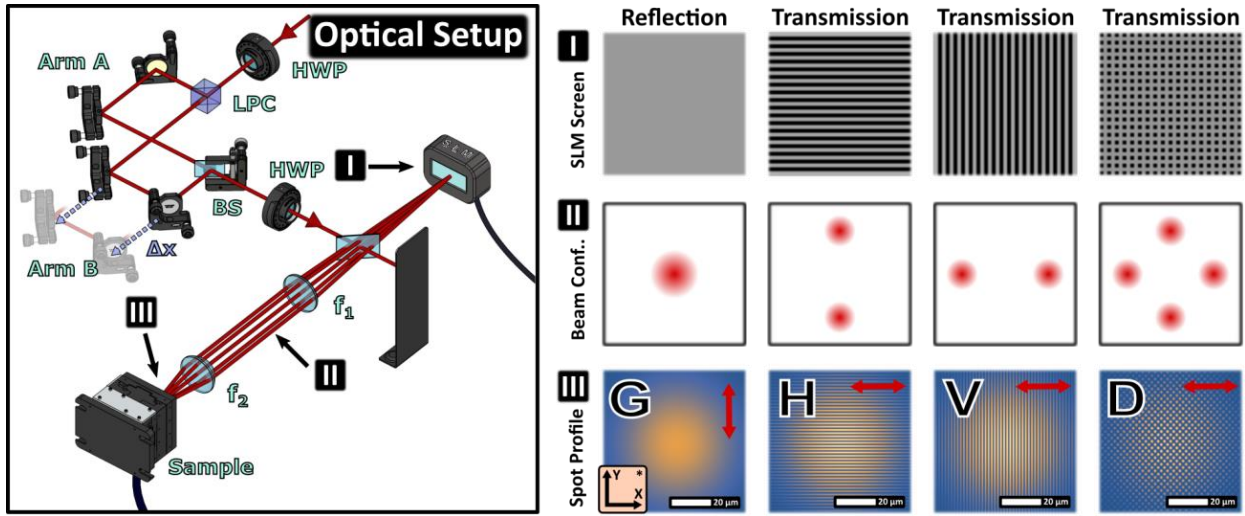

Figure 1.Left: Experimental Setup. Abbreviations: Half-Wave Plate (HWP), Linear Polarizing Cube (LPC), Beam Splitter (BS), Spatial Light Modulator (SLM), focusing lenses ($f_1$,$f_2$); Right: SLM function (I); Spot profile distribution at the sample: Gaussian (*G*) and DLIP (*V, H, D*) profiles. The red arrow indicates the polarisation vector (III).

**b.    Theoretical model**

*DLIP Intensity profiles*

DLIP is based on the superposition of the electric field vectors of coherent beams. The electric field of a light wave *n* is equal to $\vec{E}_n = A_n \vec{p}_n e^{i(\omega t - \vec{k}_n \cdot \vec{r})}$ where $A_n$ is the amplitude, $\vec{p}_n$ is the polarisation vector, $\vec{k}_n$ (i.e. $|\vec{k}_n| \equiv k = \frac{2\pi}{\lambda_L}$) stands for the wavevector, $\vec{r}$ is the position vector, $\omega$ is the angular frequency, and *t* is the time. It is noted that the resulting interference pattern for *N* beams is calculated through the expression $\vec{E}_{total} = \sum_{n=1}^{N} \vec{E}_n$. Each laser beam irradiates the material at an incident angle with the vertical axis equal to $\theta_n$ and azimuthal angle $\varphi_n$. It is assumed that all laser beams are linearly polarised and the polarisation direction is the same for all DLIP beams (i.e. perpendicular to that of the Gaussian). The total spatial intensity distribution is, then, provided by the formula $I_0 = c\varepsilon_0 |\vec{E}_{total}|^2/2$, where *c* and $\varepsilon_0$ stand for the speed of light and dielectric vacuum permittivity, respectively. For interference with *two* ($\varphi_1 = \varphi_2 = 0$, and $\theta_1 = \theta_2 = \theta$) and *four* beams ($\varphi_1 = \varphi_2 = 0$, $\varphi_3 = \varphi_4 = \pi/2$, $\theta_1 = \theta_2 = \theta_3 = \theta_4 = \theta$), the following total intensities are $I_0^{(2)}$ and $I_0^{(4)}$ on the surface of the material, respectively (by calculating the Poynting vector and by taking the average value $\langle I_0 \rangle$ over the laser period $\lambda_L/c$)

$$I_0^{(2)} = I_1[1 + cos(2kxsin\theta)] \qquad (1)$$
$$I_0^{(4)} = I_1 \left\{ \left[ 1 + \frac{1}{2}cos(2kxsin\theta) + \frac{1}{2}cos(2kysin\theta) + 2cos(kxsin\theta)cos(kysin\theta) \right] \right\}$$

, where $I_1$ is the intensity of each of the constituent laser beams of the DLIP in which a Gaussian envelope is included ($\sim e^{-2\left(\frac{x^2+y^2}{R_0^2}\right)}$) with an $e^{-2}$ Gaussian spot radius equal to $R_0$; similarly, $I_1$ includes the Gaussian distribution in time, $e^{-4ln(2)\left(\frac{(t-3\tau_p)^2}{\tau_p^2}\right)}$. It is noted that a Cartesian coordinate system is used (*x,y,z*). It is evident that the choice of $\theta_n$ can be used to define the periodicities of the interference pattern. More specifically, for two-beam DLIP, a sinusoidal (of periodicity equal to $\Lambda_{DLIP} = \frac{\lambda_L}{(2sin\theta)}$ along *x*-axis) or for a four-beam DLIP, a dot-type intensity distribution (of periodicity equal to $\Lambda_{DLIP} = \frac{\lambda_L}{(sin\theta)}$ along *x*-axis and *y*-axis and a secondary $\Lambda_{DLIP} = \frac{1}{\sqrt{2}} \frac{\lambda_L}{(sin\theta)}$ along the direction defined by $\hat{x}+\hat{y}$ ($\hat{x}, \hat{y}$ are unit vectors) are derived as illustrated in Figure 1.

*Modelling of laser-matter interaction*

To simulate the surface modification processes and reveal the impact of excitation levels and hydrodynamical effects following irradiation with fs pulses, a multiscale description of the underlying physical mechanisms is required to account for the response of the material. Given that the intensity profile changes spatially with the combination of temporarily separated Gaussian and DLIP pulses, a detailed investigation of the excitation, relaxation processes and propagation of the produced hydrothermal waves can be performed through a theoretical framework that comprises modules that model the processes in various temporal regimes. A detailed description of the theoretical model that simulates the physical mechanisms that characterize irradiation of stainless steel with DLIP pulses is presented in Ref. [32]. In that report, the thermophysical properties and optical parameters of the material are also provided.

In summary, a three-dimensional Two Temperature Model (TTM) represents the standard theoretical framework to investigate laser-matter interaction and the energy transfer between the electron and lattice subsystems

$$C_e \frac{\partial T_e}{\partial t} = \vec{\nabla}(k_e \vec{\nabla} T_e) - g(T_e - T_L) + W$$
$$C_L \frac{\partial T_L}{\partial t} = \vec{\nabla}(k_L \vec{\nabla} T_L) + g(T_e - T_L)$$
(2)

where $C_e$ and $C_L$ are the heat capacities of the electron and lattice subsystems, respectively while $T_e$ and $T_L$ are the temperatures of the two systems. By contrast, $k_e$ ($k_L \sim 0.01\, k_e$) stand for the electron (lattice) conductivity, $g$ is the electron-phonon coupling parameter while $W$ corresponds to the absorbed laser power density that is related to the laser intensity. It is noted that the wavelength of all laser beams is equal to $\lambda_L$=1026 nm. As the laser intensity comprises two (temporarily separated with a delay $\Delta\tau$) pulses, $W$ includes the variation of the total intensity of the laser beam (i.e. $\alpha I_{total}$, where $\alpha$ is the absorption coefficient of the material at $\lambda_L$=1026 nm). Due to the presence of two pulses, one Gaussian and one DLIP (with either two or four beams), the total laser intensity on the material surface is provided from the following expression (see derivation in the Supplementary Material)

$$I_{total}(t, x, y, surface) = \left[ P_1 e^{-4\ln 2 \left(\frac{t - 3\tau_p - G_1 \Delta\tau}{\tau_p}\right)^2} + P_2 e^{-4\ln 2 \left(\frac{t - 3\tau_p - G_2 \Delta\tau}{\tau_p}\right)^2} \right]$$
(3)

, where $\tau_p$ is the pulse duration of each laser beam and $P_1$ and $P_2$ are the two temporarily separated[36] pulses with a $\Delta\tau$ sdelay. With respect to the laser beams: (i) the Gaussian beam irradiates the material first if $G_1$=1, $G_2$=0, otherwise, $G_1$=0, $G_2$=1 if the DLIP irradiation comes first, (ii) $P_1$ (or $P_2$) is the (spatial) profile of the Gaussian (or the DLIP) beam. More specifically, for the DLIP beam, $P_2 = I_0^{(2)}, I_0^{(4)}$ which depends on the number of beams that comprise the total beam. In the experiments performed in this work, the total laser energy $E_{tot}$ that is distributed in the system is split into two parts, 2/3 $E_{tot}$ are provided from the Gaussian beam while the rest is attributed to the DLIP beam.

It is noted that the values of the energy per pulse used in the experiments (i.e. 40 μJ, 60 μJ, 80 μJ) are found, by the simulations, to result to mass removal (i.e. ablation) [32]. To account for the experimental results and considering that mass removal is observed to be minimal in the experiments, all simulations have been carried out using lower energy values of the order of ~25 μJ.

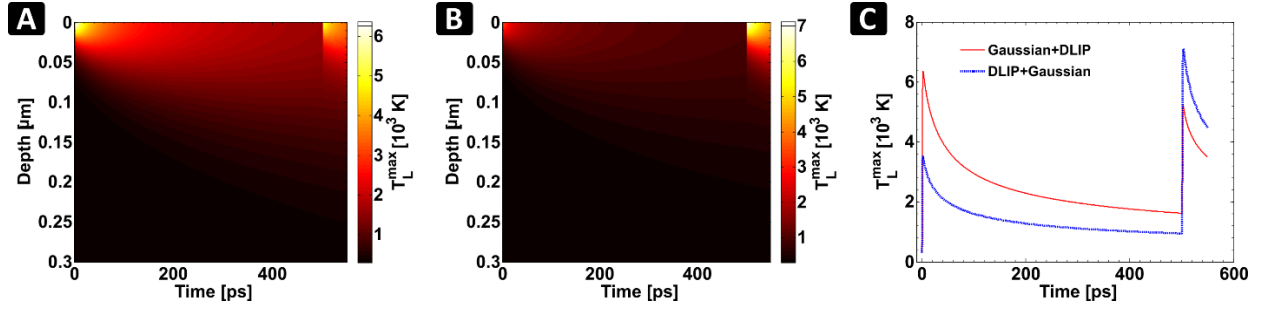

Figure 2: $T_L^{max}$ evolution at different depths for (A) $G_1$=1, $G_2$=0 and (B) $G_1$=0, $P_2$=1, after irradiating a flat profile (*NP*=1) with *E*=25 μJ, *Δτ* =500 ps.; (c) $T_L^{max}$ on the sample surface as a function of time in the cases shown in A and B.

It is emphasised that special attention is required to determine the impact of the (absorbed) laser energy given that the pulse separation in the double pulse experiment (*Δτ* =500 ps) indicates that the second pulse irradiates a material, which is in its molten phase. Results for the maximum lattice temperature $T_L^{max}$ evolution with time at various depths are illustrated in Figure 2 following irradiation of a flat surface with $E_{tot}$=25 μJ for $G_1$=1, $G_2$=0 (Figure 2A) and $G_1$=0, $G_2$=1 (Figure 2B); results show that this energy value leads to a phase change in the second case. Interestingly, when the Gaussian pulse irradiates the material first, the impact of the second pulse leads also to minimal mass removal that is simulated with the removal of material lattice points with temperatures higher than $0.9 T_{cr}$ [$T_{cr}$ stands for the thermodynamic critical temperature which for stainless steel is ~8500 K] [32, 37, 38] (Figure 2C). Although the volume of the ablated material is minimal at $E_{tot}$ =25 μJ, a substantially larger ablation is predicted at higher energies (results are not shown).

Interestingly, both the values and evolution of $T_L^{max}$ are dependent on which pulse irradiates the material first, since the energies of the two pulses are different (Figure 2C). Furthermore, the material response is related to what phase the material is in at the time of irradiation; more specifically, it is shown that if the second pulse irradiates a molten volume, a significant reflectivity variation [32] leads to a distinct thermal response of the material causing different: (i) absorption levels produced by the sequence of the double pulses and (ii) temperature gradients. The evolution of $T_L^{max}$, illustrated in Figure 2C is calculated for a flat surface. Although such $T_L^{max}$ evolutions illustrate the thermal response of the system for *NP*=1, similar conclusions are also deduced for non-flat profiles.

On the other hand, to interpret the induced surface modification, the development and evolution of hydrothermal waves and the dynamics of the produced fluid movement as a result for the phase transformation is mathematically described by the Navier-Stokes equations (NSE) [36]

$$\rho_0 \left( \frac{\partial \vec{u}}{\partial t} + \vec{u} \cdot \vec{\nabla} \vec{u} \right) = \vec{\nabla} \cdot \left( -P\mathbf{1} + \mu(\vec{\nabla}\vec{u}) + \mu(\vec{\nabla}\vec{u})^T \right) \qquad (4)$$

where $\mu$ and $\rho_0$ stand for the viscosity and density, respectively, of the molten (uncompressed) material, while $P$ and $\vec{u}$ are the pressure and velocity of the fluid and superscript $T$ denotes the transpose of the vector $\vec{\nabla}\vec{u}$, and **1** is a 3×3 identity matrix. It is noted that velocity gradient tensor is a 3×3 matrix $(\vec{\nabla}\vec{u})_{ij} = \frac{\partial u_i}{\partial x_j}$ (see Ref [36] and Supplementary Material) where $(u,v,w)$ are the components of $\vec{u}$ in Cartesian coordinates $(x,y,z)$ The solution of NSE is conducted through the employment of appropriate thermocapillary boundary conditions. More specifically, to describe the Marangoni flow, a shear stress balance on the free surface is described by the equation $\frac{\partial u_\tau}{\partial n} = -\frac{1}{\mu}\frac{\partial \sigma}{\partial T_L}\frac{\partial T_L}{\partial \tau}$ at the liquid free surface where $u_T$ corresponds to the component of the velocity on the surface tangent direction $\tau$ while $n$ indicates the component normal to the free surface. This expression turns into two equations for $NP=1$ (i.e. flat surface): $\frac{\partial u}{\partial z} = -\frac{1}{\mu}\frac{\partial \sigma}{\partial T_L}\frac{\partial T_L}{\partial x}$ and $\frac{\partial v}{\partial z} = -\frac{1}{\mu}\frac{\partial \sigma}{\partial T_L}\frac{\partial T_L}{\partial y}$. In Eq.4, $\sigma$ stands for the surface tension of the material (see [32] for parameter values and for a more detailed description of the fluid dynamics module). As discussed in previous reports [14, 32, 39], inhomogeneous energy deposition leads to the development of Marangoni effects (i.e. surface tension-driven molten material flow) and displacement of material from regions of high to low temperatures. In addition, it is assumed that non-slipping conditions ($\vec{u} = 0$) are applied on the interface between solid- and liquid interface. Finally, the molten material is considered to be an incompressible fluid ($\vec{\nabla} \cdot \vec{u} = 0$). To simulate the fluid movement and dynamics of the molten material, the affected region has been divided in two subregions, one that contains a material in solid phase and another in liquid phase. The hydrodynamic equations are solved in both regions. To include the 'hydrodynamic' effect of the solid domain, material in the solid phase is modelled as an extremely viscous liquid ($\mu_{solid}=10^5\ \mu$), which will result into velocity fields that are infinitesimally small. An apparent viscosity is then defined with a smooth switch function of Gaussian form to emulate the step of viscosity at the melting temperature [14]. A moving boundary for the two phases (solid-liquid interface) is based on position of the isothermal $T_L=T_{melting}$, where $T_{melting}$ stands for the melting point of the material.

As the theoretical model aims to describe both the thermal response and dynamics of the molten material that leads eventually to morphological changes, topography variations should be considered due to consecutive irradiation as a result of increasing the energy dose. Therefore, apart from the aforementioned boundary conditions used to solve NSE (i.e. shear stress balance on a free but nonflat surface), appropriate conditions need to be introduced for the electron and lattice temperatures. More specifically, negligible heat loss from the free surface should be considered while it is also assumed that at large distances from the affected region both the electron and lattice temperatures are ~300 K. Special attention is required to evaluate the lattice temperature resulting from multiple irradiations. While the initial conditions for the velocity and lattice/electron temperatures before the delayed pulse irradiates the material are determined from the combined solution of NSE and TTM at the moment the second pulse heats the material, it is assumed that every consecutive double pulse in the train always irradiates a material at room temperature.

## Results and discussion

### A. Single pulse irradiation

The morphology obtained following irradiation with trains of single pulses are illustrated in Figure 3, assuming various intensity profile types: Gaussian distribution (*G*), vertical DLIP (*V*), horizontal DLIP (*H*) and a DLIP profile produced with four beams (*D*) (Figure 3A). Different pulse energies were used for the Gaussian and the DLIP distribution in order to maintain similar effective fluence value on the irradiated areas; the energy per pulse is 80 μJ for *G*, and 57 μJ for *V*, *H* and *D*. The total number of pulses is *NP* = 50 in all cases. For the Gaussian distribution subwavelength LSFL are formed throughout the irradiated area with orientation perpendicular to the laser beam polarisation (Figure 3B) as reported in previous works [1, 14, 40]. For the vertical DLIP some small corrugation has been observed in-between the DLIP pattern (Figure 3, V). The type of the additional corrugation is dependent on the DLIP periodicity. More specifically, small protrusions inside the crater are produced if the DLIP periodicity is comparable to $\lambda_L$, otherwise, LIPSS are formed for $\Lambda_{DLIP} > \sim 2\, \lambda_L$ [32]. The formation of these structures have been reported in a previous work [32]. On the other hand, for the horizontal DLIP, LIPSS are formed unperturbed among the DLIP grooves (Figure 3, H). It is possible that a surface plasmon wave-assisted mechanism is capable to explain the fabrication mechanism of those structures, however, further investigation is required [27]. Finally, for a four beam-based DLIP, formation of ripples inside holes produced in regions of high intensity is observed [32] (Figure 3, D).

The FFT diagrams corresponding to the aforementioned topographies are illustrated in Figure 3C. In these FFT images, the spatial frequency values of LSFL structures correspond to the green dotted circles, while those of DLIP structures to the blue ones. More specifically, for *G*, the calculated LIPSS periodicity is equal to ~818 ± 7 nm. On the other hand, for *V*, the competition between DLIP and the impact of *G* leads to a more complex profile that is indicated by the presence of an additional protrusion corresponding to a larger periodicity As a result, the period of the DLIP, for *V*, is $\Lambda_{DLIP} = 1660 \pm 4$ nm marked with blue circles in the corresponding FFT diagram. In the case of

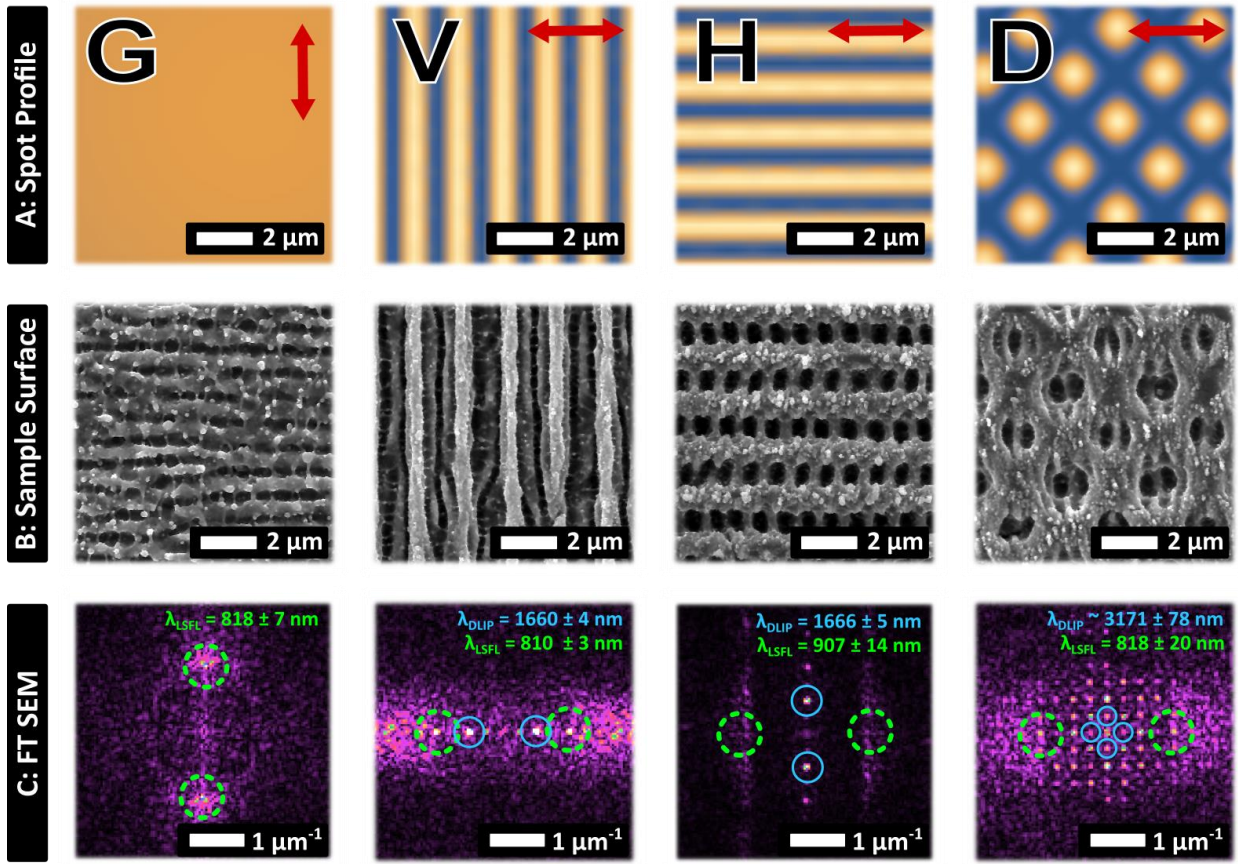

Figure 3: A: Spot intensity distribution; B: SEM images of processed surface; C: Fourier transform diagrams, produced by trains of single pulses. The total number of pulses is *NP* = 50, the energy per pulse is $E_{tot}$ = 80 μJ for *G*, and $E_{tot}$ = 57 μJ for the DLIP distribution *V*, *H* and *D*. The red arrow in A indicates the polarisation direction.

*H*, the LSFL formation is not disturbed by the DLIP pattern due to the configuration of the polarisation with the DLIP. Results indicate that the induced LSFL have a period of 907 ± 15 nm for $\Lambda_{DLIP}$ = 1666 ± 5 nm. The areas corresponding to LSFL frequencies are marked in the FFT diagram, nonetheless the zone does not feature a clear peak, compared to the other cases, due to the fact that ripples formation occurs only inside the DLIP groove. In the case of *D*, the period obtained from the FFT diagram in the horizontal and vertical axes are $\Lambda_{DLIP}$ = 3230 ± 100 nm and $\Lambda_{DLIP}$ = 3114 ± 60 nm, respectively, which is equivalent to the distance between the craters. Finally, the period of the LIPSS formed inside the craters is measured to be equal to 818 ± 20 nm, (i.e. similar to that of the *G* beam).

## B. Combining a Gaussian with a DLIP pulse in DPI

In this section the result of irradiation of stainless steel with two subsequent pulses, each having a Gaussian or DLIP profile is investigated both experimentally and theoretically. The experimental results together with theoretical simulations are shown in Figure 4, where the order of the pulses is varied as indicated. The interpulse delay was always $\Delta\tau$ = 500 ps, while the total energy per pair of pulses was $E_{tot}$ = 80 μJ and *NP* = 50. The label of each figure indicates the type of the distribution and the order of the pulses; for example, "*G*↕+*D*↔" in Figure 4E corresponds to a morphology that was obtained after irradiation of the sample with 50 double pulses where the Gaussian pulse precedes while the second 2D DLIP pulse irradiates the material after 500 ps. To account for the induced topography, a parametric study has been performed to describe both the interpulse and the intrapulse surface modification and how a corrugated profile influences the energy absorption and thermal response of the irradiated material [14, 32]. Theoretical predictions for the induced structures are derived in all cases and compared with the experimental findings for the laser conditions used. At the same time, appropriate experimental protocols were applied, aimed to emphasise the role of hydrodynamic melt flow due to the cumulative action of the two pulses. All the results obtained for double pulses at the different sequences tested, i.e. *G*+*V*, *V*+*G*, *G*+*H*, *H*+*G*, *G*+*D*, *D*+*G* are presented in Figure 4 and discussed below. It should be noted that the size of the arrows in the third and fourth columns of Figure 4 present an estimate of the magnitude of the speed of the material flow at each spatial point; these results are derived from the solution of the Navier-Stokes equations providing the dynamics of the molten material at those time points and constitute a guide to eye. A more detailed and accurate representation is presented in Section D, in which the spatial distribution of the magnitude and direction of the calculated fluid velocity vectors are illustrated for a representative, *G*+*V*, case. Simulation results for the rest of the irradiation schemes applied are provided in the supplementary material. Surface patterns illustrated in the fifth column of Fig.4 depict contour plots that indicate the depth variation for *NP* = 50 (see also Supplementary Material). The maximum depth attained for the pattern is ~230 nm.

(i) *G+V and V+G*

Experimental results illustrate (Figure 4) that the *G*+*V* combination leads to the formation of LSFL structures that are affected from the impact of the DLIP pulse on the irradiated area. In particular, LIPSS originating from *G*, with $\Lambda_{LIPSS}$ = 946 ± 17 nm, and DLIP structures from the *V*, with $\Lambda_{DLIP}$ = 1700 ± 55 nm, pulse smoothly overlap. Theoretical predictions show that when the *V* pulse (i.e. $\Lambda_{DLIP}$ = 1700 nm, for $\theta$=18⁰) irradiates the material after $\Delta\tau$, LIPSS that have been produced from the Gaussian pulse *G* are not destroyed from *V* in the region where the DLIP intensity is low; by contrast, shallow ripples are formed along the region where the intensity of *V* is higher. To explain the produced topography, a theoretical investigation is conducted to evaluate the role of electrodynamics, thermal effects and fluid dynamics that yield those particular morphological features. As the increase of *NP* affects the morphology, a feedback mechanism is incorporated into the model to derive the absorbed energy and response of the material. To evaluate the influence of the second beam (Figure 4), the lattice temperature on the transverse plane (i.e. *x-y* plane) is illustrated at time *t*=490 ps and at *t*=520 ps (third and fourth column images in Figure 4A, respectively), for *NP*=50. Simulations indicate that there are regions in which sufficiently high values of laser energy is absorbed along the wells of the rippled zone that leads to phase transition and therefore, *V* irradiates predominantly a molten material. The Gaussian beam dictates the development of the horizontally orientated ripples through the excitation of surface plasmons and the frequency of LIPSS is determined by the effect of the interference of SP with the incident beam [41]. For *NP*=50, the LIPSS periodicity is calculated to be equal to $\Lambda_{LIPSS}$~915 nm that is comparable to the experimental value $\Lambda_{LIPSS}^{exp}$=946 ± 17 nm. What is also observed is that relaxation processes and temperature gradients of the induced molten material lead to melt movement away from the valleys of the rippled zone and always along the (i.e. at *t*=490 ps). On the other hand, the *V* pulse increases the temperature at locations on the topography where the energy deposition from the DLIP is higher (i.e. at *t*=520 ps). As a result of the enhanced temperature gradient along the *x*-axis, melt movement occurs, indicated by respective vectors in the temperature profiles presented in Figure 4B. The combined effect leads to the formation of depressed zones at the valleys of the ripples (depicted by 'darker regions' of high temperatures in Figure 4) and to a periodic topography that agrees with the experimental results (see left column in Figure 4)). A more precise illustration of the hydrothermal melt movement at each point in a particular region that illustrates the effect of the hydrodynamical process is presented in the Supplementary material. In a *V*+*G* scheme, the DLIP ($\Lambda_{DLIP} = 1700$ nm) first irradiates the material surface; as a result, in locations where the DLIP intensity is higher, the enhanced energy absorption influences the response of the material when the Gaussian beam arrives with a delay of 500 ps. In that case, the main pattern observed is the DLIP with period of $\Lambda_{DLIP} = 1701 ± 9$ nm. Moreover, on top of the DLIP grooves, dots that originate from a spatially modulated periodic energy absorption, induced from the Gaussian beam, lead to 'dot-like' periodic structures of period equal to $\Lambda_{LIPPS} = 885 ± 32$ nm. More specifically, *G* first excites charge carriers and subsequently, through relaxation, the temperature rises in the regions among the DLIP intensity maxima, i.e. the DLIP valleys, as surface depression will take place there due to hydrothermal melt movement. At the same time, the induced surface plasmon excitation in the region outside such valleys and coupling with the incident radiation leads again to inhomogeneous energy deposition.

As a result of the temperature gradients induced within the molten material the melt is pushed away from the DLIP valleys along the *x*-axis (shown for *t*=490 ps in Figure 4B). On the other hand, the *G* pulse increases the temperature outside the DLIP valleys driving the produced hydrothermal waves along the *y*-axis (shown for *t*=520 ps in Figure 4B). The combined effect of the above phenomena gives rise to the formation of surface patterns that comply well with the experimental observations (Figure 4B). The theoretical prediction of the period of LIPSS is estimated to be ~890 nm that is similar to the experimentally measured value of ~885 ± 32 nm.

To emphasise the special role of hydrothermal movement and provide a more detailed picture of the impact of the induced hydrothermal waves, the melt fluid velocity vectors at every lattice point in a region ~3x3 μm$^2$ are illustrated for *G+V* at 490 ps and 520 ps respectively (Figure 4). These vectors present an estimate of the magnitude of the speed of the material flow at each spatial point, providing the dynamics of the molten material at those time points.

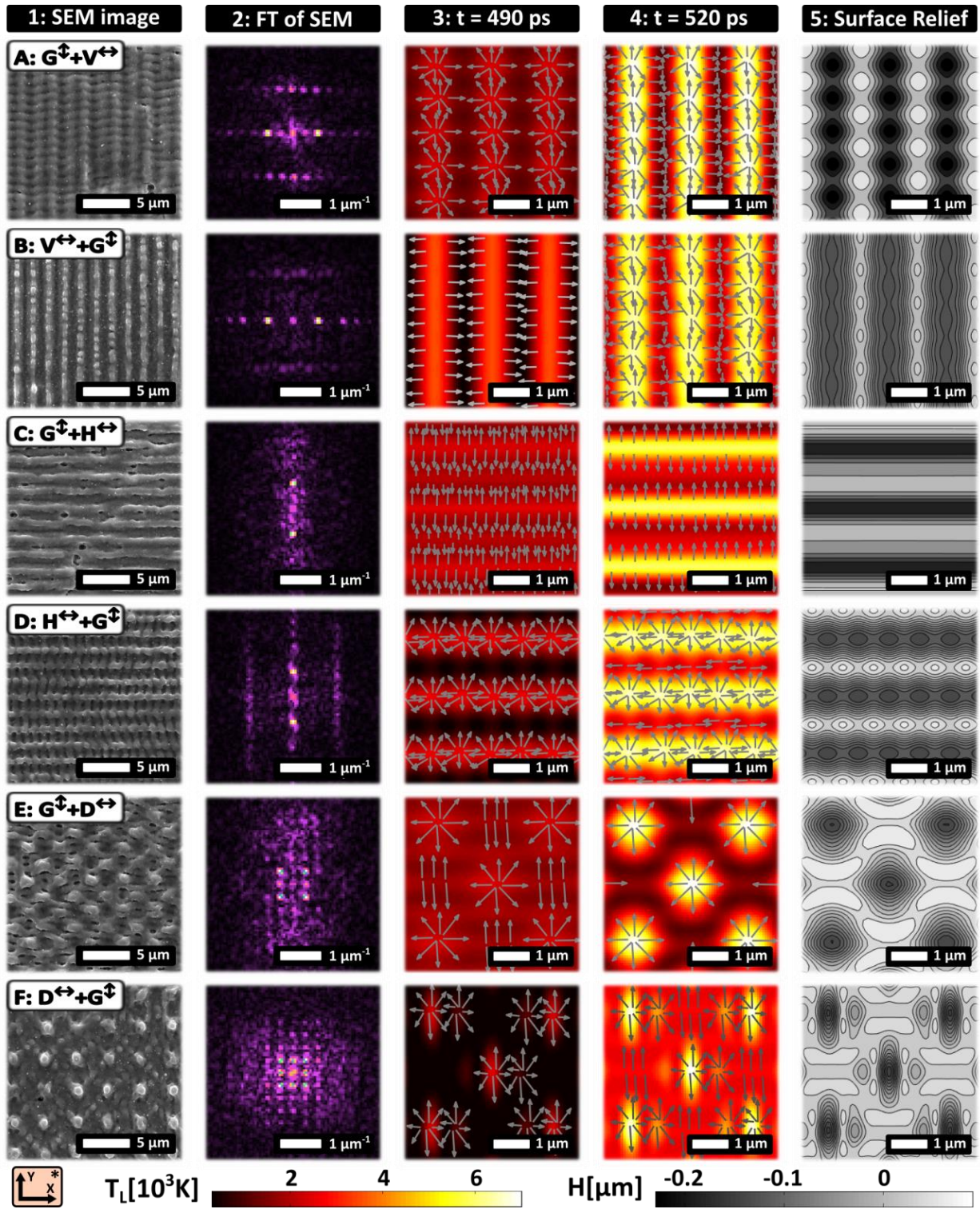

Figure 4: SEM images of the double pulse irradiation schemes used (1st column) together with the respective FFT analyses (2nd column) and simulation results (columns 3-5). The pulse profile and the order are indicated in the legends of the 1st column, while the superscript arrows indicate the pulse polarization in each case. The total energy of the double pulse pair is 80 μJ/pulse and $NP$ = 50 pulses. Simulation results are shown in each case, just before (3rd column, $\Delta\tau$ = 490 ps) and right after (4th column, $\Delta\tau$ = 520 ps) the arrival of the second pulse. The final morphology predicted by the simulation experiment is shown in the 5th column (contour plots to illustrate the height). The arrows in the 3rd and 4th column provide a guide to eye representation of the fluid transport direction. The color bars at the bottom of the figure provide the temperature (3rd and 4th columns) and height (5th column) values range.

(ii) *G+H and H+G*

In the case of a double pulse experiment in which the contributing pulses to the DLIP yield an interference pattern parallel to the *x*-axis, the impact of the DLIP pulse leads to a different surface topography. More specifically, as in the *G+V* case, it is expected that the first pulse (*G*) excites surface plasmons and leads to subwavelength LIPSS oriented perpendicularly to the laser polarisation that will lead to a rippled profile. Nonetheless, analysis of the experimental results indicate that only one ripple frequency is observed, that of the DLIP, with a period of $\Lambda_{DLIP}$ = 1630 ± 14 nm. In Figure 4C, the respective temperature profile at *t*=490 ps is illustrated, i.e. before the arrival of the *H* pulse onto the hot material. Upon the incidence of the H pulse (i.e. *t*=520 ps in Figure 4C), the temperature in the valleys of the periodic regions formed by *G* will further rise above that attained with the initial DLIP pulse. As shown by the respective simulations shown in the same figure, hydrodynamical movement drives the melt flow parallel to the *y*-axis before solidification, giving rise to a topography formation that closely resembles the experimental one. It should be noted that $\Lambda_{DLIP}$ is a multiple of the predicted LIPSS period ($\Lambda_{LIPSS}$~850 nm) and therefore there is no overlap between the patterns produced from the combined effect of *G* and *H*.

A different profile is induced for the *H+G* pulse sequence (Figure 4D). In that case, the DLIP pulse leads to the formation of long stripes (tops and valleys) periodically situated at distances defined by $\Lambda_{DLIP}$, however, the polarisation beam direction of the DLIP pulse (along the *x*-axis) yields excitation of surface plasmon waves of ~880 nm periodicity and hydrothermal waves that propagate along the *x*-axis. Similar structures have been observed and interpreted in a previous report [32]. On the other hand, irradiation with a *G* pulse with a 500 ps delay, enable excitation of surface plasmon along the *y*-axis, with hydrothermal waves propagating along this axis, as well and yielding LIPSS parallel to the *x*-axis. As noted in the previous paragraph, the fact that the $\Lambda_{DLIP}$ is a multiple of the predicted period of the LIPSS, produced from *G*, leads to surface patterns similar to experimental results (Figure 4D). By contrast, a different relation between the periodicity due to the $\Lambda_{DLIP}$ and *G* would lead to a distinctly different profile with a substantial overlapping of the respective periodicities. In Figure 4D, the temperature profile and melt movement are illustrated at *t*=490 ps and *t*=520 ps, respectively. According to the simulation results, vertical ripples produced by *H* are much more pronounced in *H+G* case because the melt flow is already developed upon arrival of *G*, in contrast to the *G+H* pulse sequence case.

(iii) *G+D and D+G*

Interesting morphologies are obtained when a combined *G* plus a *D* pulse with $\Lambda_{DLIP}$ =3100 nm is used (Figure 4E). Theoretical results indicate that the action of the *G* beam leads to an expected surface plasmon excitation and hydrothermal waves, while the *D* pulse increases further the temperature at points where the DLIP intensity is high. Thus, a periodic profile is produced ($\Lambda_{LIPSS} = 824 \pm 46$ nm) with the ripples oriented parallel to the *x*-axis that are interrupted with periodically situated subsided regions due to the impact of the *D* pulse (Figure 4E). The latter regions exhibit the ripple periodicity attained by *G*. The respective simulated temperature profile as well as the propagation of the induced hydrothermal waves are presented in Figure 4D for $t$=490 ps and $t$=520 ps respectively, for *NP=50*.

By contrast, when *D* precedes the irradiation to G, a quite different profile, comprising high-aspect ratio asperities, is obtained. The horizontal spacing among those structures is 3163 ± 55 nm, while the vertical one is 3072 ± 67 nm, coinciding with the distance between the craters in the *D* case of Figure 3. Among the peaks, traces of LIPSS with a period in the range of 620-720 nm are observed both in horizontal as well as in vertical direction. More specifically, the intensity profile of *D* (Figure 2) leads to a temperature rise in the locations where the energy deposition is high. Furthermore, repetitive irradiation with *D* (at *NP*>2) will enable excitation of surface plasmons and the subsequent formation of ripples oriented parallel to *y*-axis. Following irradiation with the *G* pulse, i.e 500 ps after the impact of the first pulse, the material in the region which is not affected by the *D* pulse will be excited and surface plasmon waves, before leading to a rippled periodic pattern similar to the one observed in SEM images (Figure 4F). The respective temperature profile and hydrothermal melt movement at $t$=490 ps and $t$=520 ps, respectively, are illustrated in Figure 4F.

To summarise, the experimental observations and simulations indicate the following: comparing *G+V* and *V+G*, LIPSS originating from *G* are more pronounced when *G* comes first. Comparing *G+H* and *H+G*, LIPSS resulting from the action of *H* are observed only in the case that *H* irradiates the material first. Similarly, for *G+D* and *D+G*, LIPSS are more pronounced when the *G* precedes the action of *D*. LIPSS are always formed perpendicular to the polarisation of the first pulse. In addition, the above discussion shows that promotion of SP excitation-based mechanisms for the formation of LIPSS are influenced a lot on the following: (i) *G* always leads to excitation of SP waves, (ii) the DLIP polarisation and periodicity (as also shown in Ref.[32])account on whether SP excitation is possible or fluid transport will determine surface pattern formation, (iii) The order of irradiation with Gaussian and DLIP pulses can impact on the conditions for periodic structure formation through SP excitation or fluid movement along prepatterned topographies.

### C. Generation of complex patterns

In the previous sections, the effect of different pulse intensity profiles and pulse sequences was investigated and evaluated. It is evident that the spatio-temporal intensity distribution significantly influences the hydrothermal melt waves and the subsequent superposition of fluid transport derived from each pulse separately. It was shown that the combined effect of different intensity profiles predominantly influences the behaviour of the produced molten material through variation of its vorticity; this effect has proven to influence the final topography attained, giving rise to patterns of higher geometrical complexity. Two parameters that are expected to tune the surface morphology are the laser energy and the number of pulses; the increase of both parameters leads to higher temperature gradients and therefore to large vorticity changes. A characteristic example of the evolution of surface topography upon increasing *NP* at specific pulse energy is presented in Figure 5 (rows), for the *D+G* pulse sequence.

Results demonstrate the pronounced influence of the first pulse in the sequence in the feedback mechanism, while each subsequent dose of the series enhances the depth of the induced topography. Similar results (not shown here) are derived for the rest of the pulse sequences. On top of that, the role of hydrothermal phenomena and the impact of the topography in the produced temperature gradients that lead to a dynamic evolution of the surface patterns at increasing dose, are very critical. Simulation results indicate that deeper profiles promote larger energy deposition locally that influences the melt movement.

In Figure 5 indicative results on the combined effect of pulse energy and energy dose (*NP)* on the induced topography, upon irradiation with the *D+G* sequence, are also shown. Notably, significant variations in the surface morphology attained, is observed, and an ensemble of complex structures can be realized. In particular, for $E_{tot}$ = 80 µJ and for low *NP* random LIPSS structures are formed, while the fingerprint of the DLIP pattern becomes evident at elevated *NP*. Similarly, for $E_{tot}$ = 60 µJ and *NP* = 10, the DLIP undulation is not visible and the surface comprised LIPSS patterns oriented in different directions. For low *NP*, neighbouring ripples create square patterns, while at the higher *NP*=20 a supra-wavelength morphology resulting from DLIP takes place. Further irradiation (e.g. *NP*=50) reveals the impact of the DLIP, as the DLIP pattern is clearly visible on the surface surrounded by dotted structures (see also Figure 5). Finally, at even lower energies ($E_{tot}$ = 40 µJ and *NP* = 50), LIPSS-like complex structures, resulting from the combined effect of the two pulses, become present. These results indicate that hydrothermal waves possibly coupled with mass removal effects should be the predominant factors that explain the induced topography. Comparing the three images of high magnification (Figure 5, zoomed region for *NP*=50) the impact of fluence in the generation of high-complexity structures, is unveiled. When the fluence is low and close the ripple formation threshold ($E_{tot}$ = 40 µJ) ripples are only formed in perpendicular directions due to DPI. When the fluence is higher ($E_{tot}$ = 60 µJ) the temperature gradient applied by the DLIP pattern is strong enough to displace material and form valleys. Blue arrows indicate the trajectory of the flow and the orange dotted circle the area that the material is accumulated. In even higher pulse energy ($E_{tot}$ = 80 µJ) the temperature gradient is stronger, and the volume of the displaced material is higher (blue arrows in Figure 5, zoomed region for *NP*=50 and at 80 µJ). In that case the displaced material is accumulated away from the areas irradiated with higher intensity and forms the observed protrusions marked by the orange dotted circle. Simulations have been performed for $E_{tot}$ = 20 µJ to illustrate the topographies attained at low energies and the role of hydrothermal phenomena. Results show that the ultrafast dynamics and the induced hydrothermal effects lead to a different surface profile than the one produced for 25 µJ (shown in 4). More specifically, lower energies produce smaller temperature gradients therefore the fingerprint of either *D* or *G* pulses will be more pronounced.

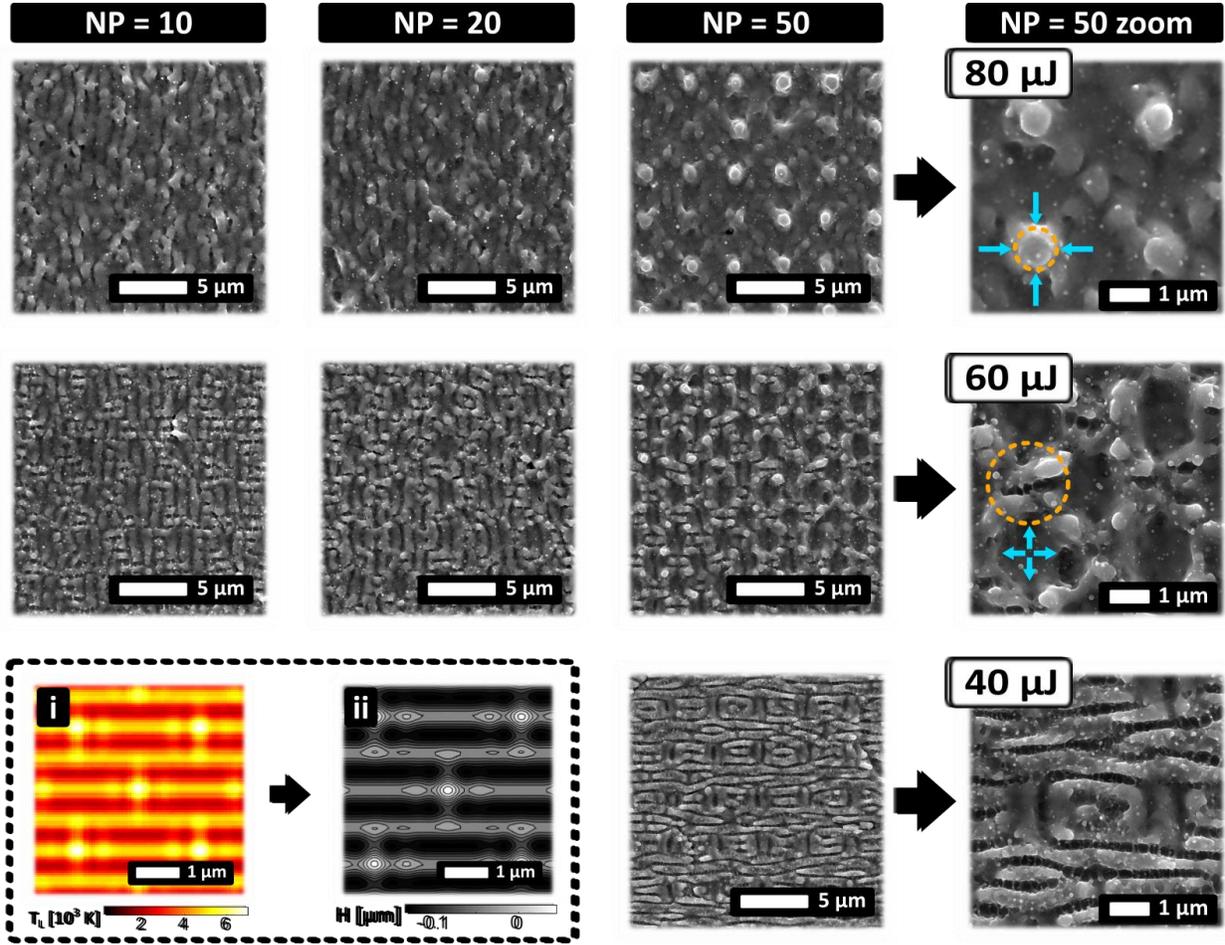

Figure 5: Stainless steel surface morphology following variation in pulse energy and *NP*, as indicated in the respective legends, for the *D+G* pulse sequence case. The corresponding result of theoretical simulations for 20 μJ pulse energy is shown as an inset defined by the dashed rectangle, illustrating (i) the temperature profile for *NP* = 50 and *t* = 520 ps and (ii) the final surface morphology for *NP* = 50.

The ability to simultaneously control the two key structure formation mechanisms, namely the inhomogeneous absorption and the microfluidic melt motion, offers potentially endless possibilities in controlling laser induced morphology in the submicron scale. Apart from the complex structures presented above, in the following we show that suitable combination of DPI parameters enable the formation of hierarchical complex structures as well, which in some cases are similar to those observed in natural systems [42]).

For example, Figure 6 presents the generation of a multiple-length-scale morphology upon irradiation with $E_{tot}$ = 70 μJ and *NP* = 100 for the *D+G* irradiation scheme. The features of the three prominent periodicities, visualized via a reciprocal FFT algorithm are presented in the same Figure (Reconstructed Surface Profile). In particular, the first with smaller period (~500 nm) corresponds to LIPSS which are oriented perpendicularly in adjacent areas, similarly to $E_{tot}$ = 40 μJ and *NP* = 50 (see Figure 5). The highest one corresponds to the effect of *D* pulse of DLIP, whilst the middle one corresponds to the 1D DLIP period, when the *V* and *H* schemes are combined. The advantage of such complex surface morphology attained, compared to the previously reported DLIP morphologies, is that it combines multiscale structures (DLIP and LIPSS) over the whole processed area in a single step.

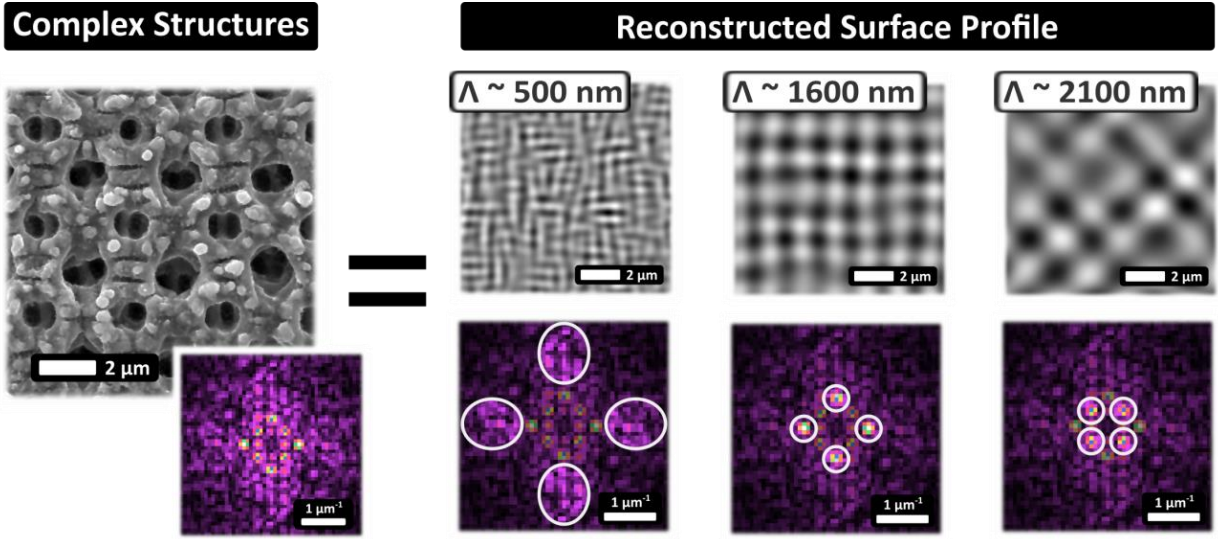

Figure 6: Complex structures generated with $E_{tot}$ =70 μJ, $NP$ = 100 with the $D+G$ pulse sequence irradiation scheme. The SEM image is analysed with FFT and reconstructed surfaces comprising structures at different length scales are illustrated. The areas corresponding to the different periodicities are highlighted in FFT diagrams.

## D. Underlying mechanism

In the previous sections, the characteristics of complex structures' formation upon irradiation of stainless steel with double pulses of a Gaussian and DLIP profiles, were investigated, both experimentally and theoretically. The impact of parameters such as the pulse profile, order of pulses, interpulse delay, pulse energy and energy dose were presented and discussed. A key element of the formation processes is the dynamic variation of the hydrothermal melt movement due to the action of the two irradiation pulses, which depends on the interpulse delay. In particular, at $t$<500 ps a melt flow is developed by hydrothermal waves created due to induced temperature gradients. Upon the impact of the second pulse the temperature profile changes drastically, as indicated by the pronounced decrease of the reflectivity in areas that are in molten phase [32]; indeed such pre-heated regions enable high energy absorption levels locally when the second pulse irradiates the material. As a consequence of the absorbed energy following the second pulse, together with the pre-existing surface temperature due to the first pulse, a complex temperature profile is generated, that is unfeasible to be created by single pulse irradiation. To further shed light on the effect of such temperature profile on hydrothermal melt movement, we present in Figure 7 simulation results, for the G+V case, for very low $NP$ values, i.e. during the first steps of structures' formation. When $NP$ = 2, the molten material at $t$ = 490 ps (Figure 7A) develops a complex flow trajectory that differs substantially from that observed at $t$= 520 ps (Figure 7B), as a response to the applied temperature profile. Considering that resolidification occurs in the order of nanoseconds, for the range of energies tested [43], a significant volume of the material has been already displaced upon the impact of the second pulse. Therefore, the overall trajectory followed by the molten material upon a single pair of double pulse irradiation can be considered as the product of the two separate trajectories imposed by each individual pulse. This demonstrates that the shape of the induced structures can be controlled by imposing different pulse profiles with suitable interpulse delays.

A key element is the value of the pulse delay (~500 ps). The choice of such a large value for the pulse separation between the two constituent pulses of the train of pulses was to illustrate the impact of the molten material when the second pulse arrives [32]. Although electron-phonon relaxation processes are completed within some tens of picoseconds and molten material is produced during that timescale, the lattice temperature of the material is quite high (>6000 K as seen Fig.3C). Furthermore, high temperature gradients (that occur till up to some hundreds of ps) is set to influence greatly the response of the material leading to complex dynamics. Thus, any effort to highlight the impact of the hydrothermal profile that is generated before the second pulse irradiates the material will be hindered. By contrast, if we allow longer pulse separations, the temperature of the fluid relaxes to substantially lower (but still above the melting point) values and therefore it is easier to reveal the variation in the fluid dynamics that is caused as a result of exposure to the second pulse.

The surface relief attained in the early stages of structures' formation will eventually determine the topology of the structures obtained for elevated number of pulses, due to the optical and hydrodynamic feedback mechanisms. Indeed, upon increasing the dose to $NP = 5$, the temperature gradient of $t= 490$ ps and $t = 520$ ps differ mainly in amplitude, but not in topology. The stability of the topology of the structures is evident in the fact that the vector fields, observed in Figure 7B and C, are quite similar. Conclusively, the combined action of the two pulses has greater impact in the first steps of the structure formation, defining the pattern of the structures, whilst the extraordinary trajectory followed by the molten material as $NP$ increases, enhances the structures' aspect ratio. Following the formation of significantly deep valleys and high hills, i.e in $NP \sim 50$ (see Supplementary Material), the surface morphology will prevail over the pulse profile in the determination of the temperature gradient. It is evident that the laser conditions and selection of $NP$ (i.e. $NP \leq 50$) were appropriately chosen to correlate experimental observations with simulations results. Certainly, higher laser fluences or increasing $NP$ are expected to produce deeper profiles and gradually alter the surface topography. This might be the objective of further investigation, however, in the current study we focused on how consecutive pulses of different intensity profiles influence the final topography. On the other hand, it is also important to correlate the value of $\Delta\tau$ with particular characteristics times such as those associated to electron-phonon relaxation and Marangoni flow. The pulse separation is much larger (about two order of magnitude) than the electron-phonon relaxation time. By contrast, it is much smaller than the expression that has been used in some reports [44, 45] as the characteristic time for Marangoni flow effect

$$\tau_m = \frac{\eta L^2}{\left|\frac{d\sigma}{dT_L}\right| T_m h} \tag{5}$$

where $\eta$ stands for the dynamic viscosity, $T_m$ is the melting temperature, $h$ corresponds to the thickness of the volume in liquid phase, $\left|\frac{d\sigma}{dT_L}\right|$ is the absolute value of the derivative of the surface tension with the lattice temperature and $L$ is a radial dimension. Assuming the values for the parameters for stainless steel, and setting $L$ (>10) μm, $h$=20 nm, $\tau_m$>50 μs. According to the use of $\tau_m$ as a criterion that determines when thermocapillary effects become competent in the surface modification process, it appears that the Marangoni flow does not appear to play the predominant role in the surface structure formation at very small timescales. The fact, though, that pulse separations of Δτ=500 ps lead to significant surface modification and simulations demonstrate the impact of thermocapillary effects indicate that further exploration of Eq.5 is required as it yields an underestimation of the fluid movement due to surface tension.

The analysis performed in this work demonstrated the impact of the polarisation-dependent process on the LSFL

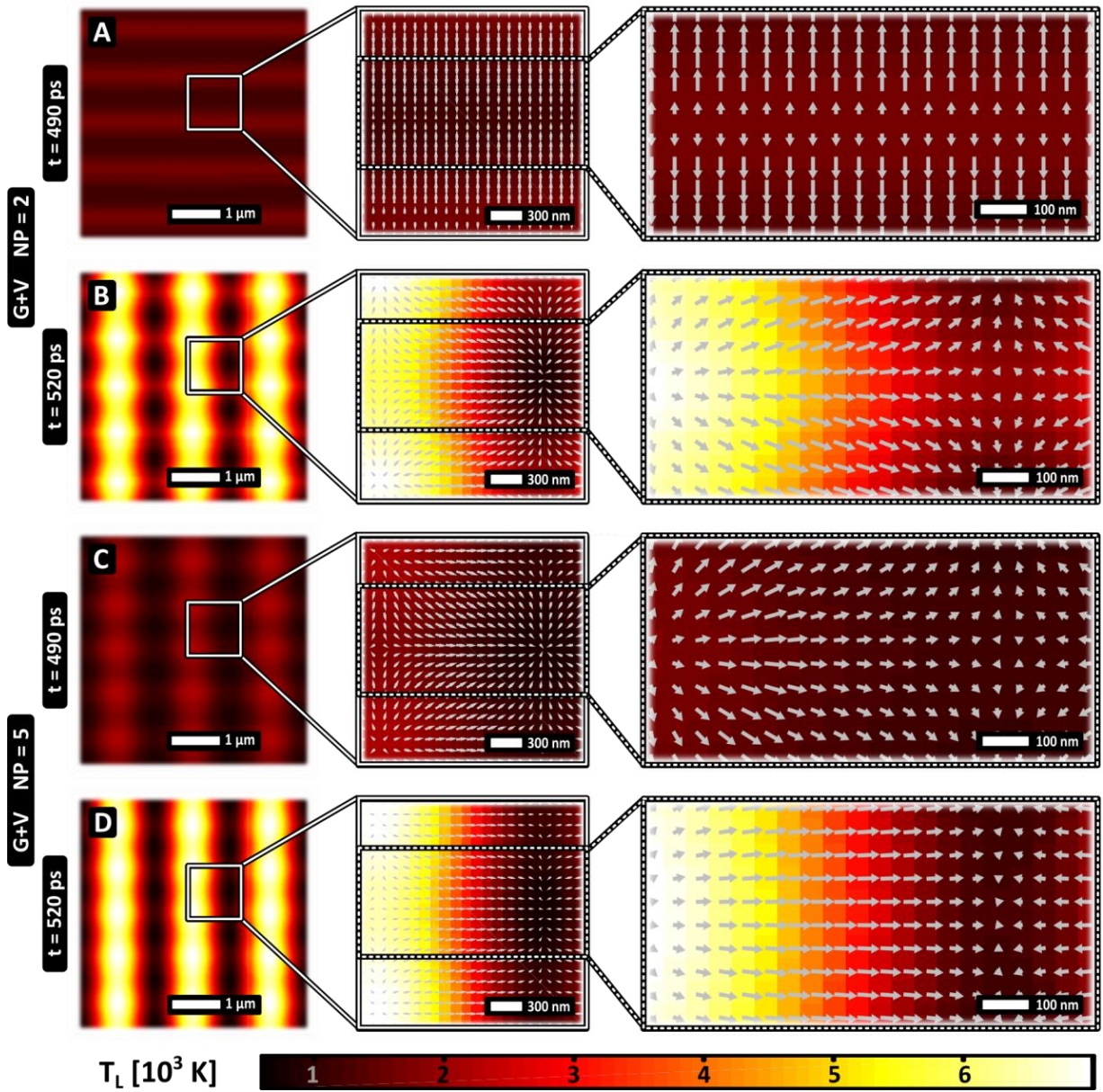

Figure 7: Simulation results showing the surface temperature profile obtained for the *G+V* case at different time moments and *NP*, as indicated. The arrows in the figures in the zoomed regions illustrate the spatial distribution of the magnitude and direction of the calculated fluid velocity vectors.

formation for either *G* or DLIP pulses. Experimental results showed that while periodic patterns were visible at small pulse energy levels or doses, the DLIP pattern imprint became more pronounced at higher pulse energies (Figure 5). As shown in Figure 5, experimental observations are significantly sensitive to the laser conditions and small changes in the parameters such as the energy values or dose are expected to lead to different topographies due to the involvement of more complex mechanisms (i.e. ablation processes, stronger temperature gradients, overlapping between DLIP and surface plasmon wave period, etc). Although the applied conditions used in this work were predominantly aimed to interpret the experimental observations in the absence of ablation effects (discussed previously in [32]), a generalized scenario can be addressed and investigated in more detail by incorporating appropriate modules in the theoretical model. For example, a more precise evaluation of the spatial distribution of the laser energy and the energy absorption from the material can be achieved through the employment of advanced computational electrodynamics based models [22]. Furthermore, appropriate modules based on atomistic molecular dynamics could also be incorporated to provide an alternative description of microscopic processes related to kinetics and phase transition-related mechanisms [46]. Another issue that requires further investigation is whether the large temperatures that are reached during the exposure of the material to a train of double pulses (>6000 K) are sufficient to cause ablation/evaporation. More specifically, the attained temperatures are substantially higher than the melting (1800 K)/boiling (3100 K) points. As emphasised in this work and previous reports [32, 37, 38], mass removal is associated with temperatures larger than $0.9T_{cr}$; therefore more exploration is needed to evaluate if that threshold constitutes an overestimation of evaporation/mass removal. Nevertheless, while further development of the model towards providing a more complete description of the processes could be the objective of a future work, the present approach demonstrates in a consistent way the capability to control laser-matter interaction through tailoring the features of the spatial intensity profile. The multiscale model that was employed is aimed also to allow an optimisation of the pattern features assuming the influence of the underlying physical processes and more importantly the role of hydrodynamical fluid transport. The agreement of simulation and experimental data demonstrates that using double, spatially tailored, fs pulses it is possible to actively control the melt microfluidic flow and in principle to make possible the generation of laser induced structures by design.

## Conclusions

In conclusion, the investigation performed in this work indicated that combining Gaussian beams with DLIP in double pulse trains enables the generation of unique sub-micron surface topographies with increased complexity. Results showed that the order of the Gaussian and DLIP pulses in train sequences influences the final surface profile attained. Furthermore, experimental results and multiscale modelling revealed that both the spatial intensity of the two pulses, as well as hydrodynamical effects have a significant impact on the pattern features. Our experiments and results presented here emphasise the capability to actively tailor the microfluidic melt motion that dominates the structure formation process, via controlling the applied temperature gradient's temporal profile. More important, the unique irradiation schemes examined here lead to the generation of novel complex morphologies comprising features in multiple length scales. This demonstrates an unparalleled capacity towards tailoring laser-induced morphology and obtaining complex topographies for a variety of applications.

## Acknowledgements

The authors would like to acknowledge support by the European Union's Horizon 2020 research and innovation program through the project *BioCombs4Nanofibres* (Grant Agreement No. 862016). Furthermore, the authors acknowledge A. Manousaki for the SEM images and A. Oikonomakis for the setup scheme.

# SUPPLEMENTARY MATERIAL

# ULTRASHORT PULSED LASER INDUCED COMPLEX SURFACE STRUCTURES GENERATED BY TAILORING THE MELT HYDRODYNAMICS

F. Fraggelakis [1*], G. D. Tsibidis [1,2♣] and E. Stratakis [1,2♦]

[1] *Institute of Electronic Structure and Laser (IESL), Foundation for Research and Technology (FORTH), N. Plastira 100, Vassilika Vouton, 70013, Heraklion, Crete, Greece*

[2] *Department of Physics, University of Crete, 71003 Heraklion, Greece*

## A. Temperature profiles and fluid movement

Simulations results and enlarged images that illustrate the temperature profiles and molten material transport at the timepoints $t$=490 ps (before the second of the double pulse train irradiates the material) and *t*=520 ps (after thermalisation following the impact of the second pulse) are presented below for *NP*=50. Results are classified according to the order of the pulses in the sequence (see main text for further explanation). The size of the velocity vectors that indicate the fluid movement are rescaled to the maximum lattice temperature value attained at each time point. The direction of the vectors indicate the fluid direction based on the temperature gradient (i.e. spatial change of the temperature). *White* dots that appear is some regions indicate a stagnant behaviour (i.e. nearly immobile fluid). A *blue-to-red* colorbar was used to emphasise better the range of temperature values and indicate more clearly the fluid direction.

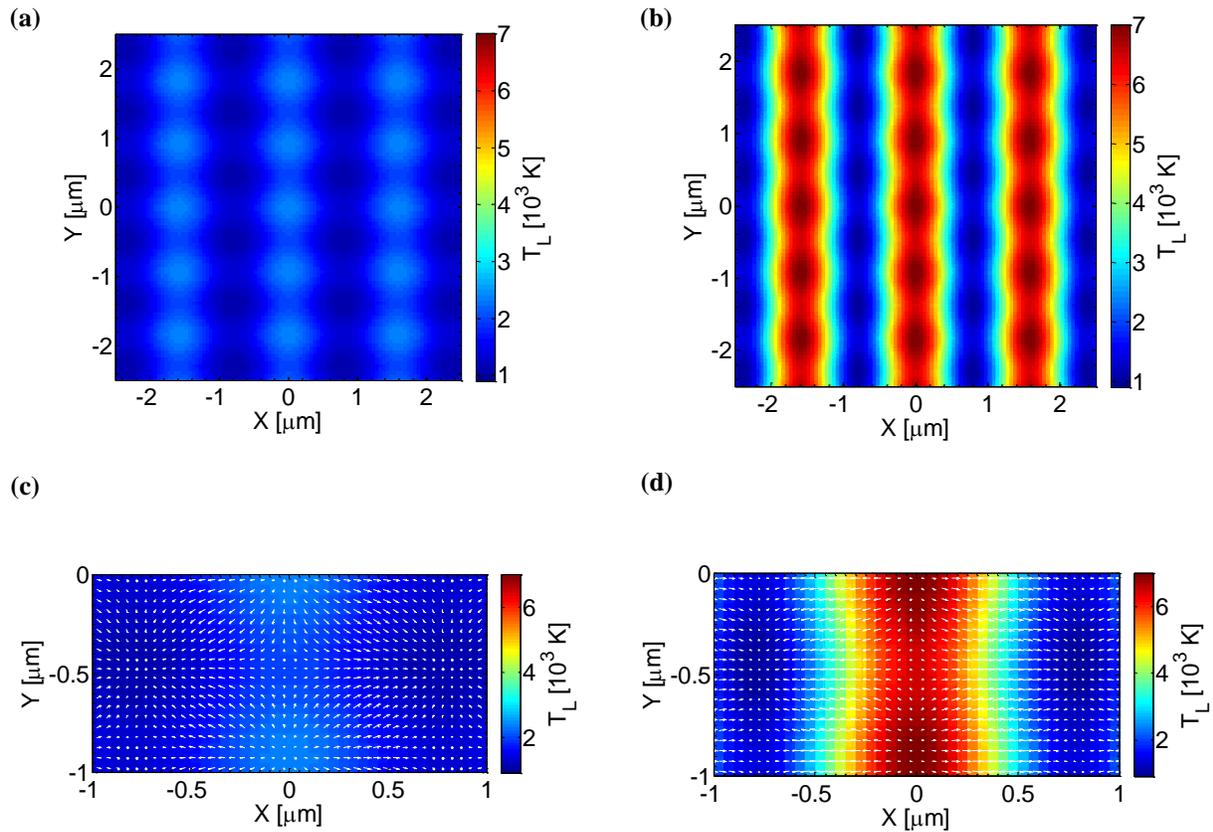

Figure 1: *G+V*: Temperature profiles at (a) *t*=490 ps (a) and (b) *t*=520 ps. Fluid movement is illustrated in (c) and (d) for *t*=490 ps and *t*=520 ps, respectively.

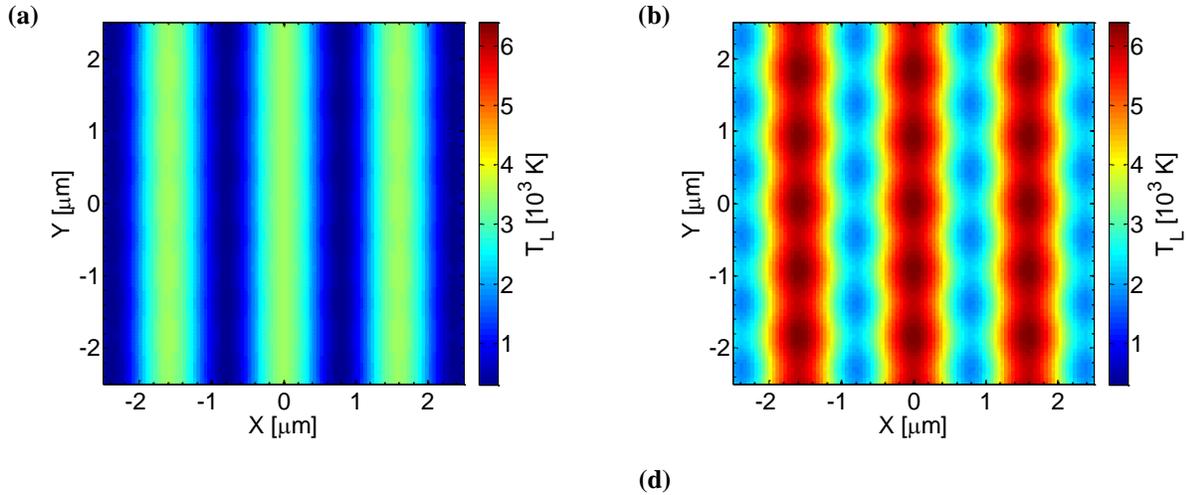

(d)

Figure 2: *V+G*: Temperature profiles at (a) *t*=490 ps (a) and (b) *t*=520 ps. Fluid movement is illustrated in (c) and (d) for *t*=490 ps and *t*=520 ps, respectively.

Figure 3: $G+H$: Temperature profiles at (a) $t$=490 ps (a) and (b) $t$=520 ps. Fluid movement is illustrated in (c) and (d) for $t$=490 ps and $t$=520 ps, respectively.

Figure 4: *H+G*: Temperature profiles at (a) $t$=490 ps (a) and (b) $t$=520 ps. Fluid movement is illustrated in (c) and (d) for $t$=490 ps and $t$=520 ps, respectively.

Figure 5: *G+D*: Temperature profiles at (a) $t$=490 ps (a) and (b) $t$=520 ps. Fluid movement is illustrated in (c) and (d) for $t$=490 ps and $t$=520 ps, respectively.

Figure 6: *D+G*: Temperature profiles at (a) *t*=490 ps (a) and (b) *t*=520 ps. Fluid movement is illustrated in (c) and (d) for *t*=490 ps and *t*=520 ps, respectively.

## B. Temperature profiles and fluid movement for *NP*=2 and *NP*=5 (for *G+V*)

To illustrate the impact of the surface topography and describe quantitatively the interpulse surface pattern changes, temperature profiles are shown for *NP*=2 (Figure 7) and *NP*=5 (Figure 8) at *t*=490 ps and *t*=520 ps for *G+V*. *White* dots that appear is some regions indicate a stagnant behaviour (i.e. nearly immobile fluid). In (c) and (d) a *blue-to-red* colorbar was used to emphasise better the range of temperature values and indicate more clearly the fluid direction. By contrast, a *red-to-white* colorbar was used in (a) and (b) (similar to the one used in the main manuscript).

Figure 7: Temperature profiles for *G+V* at (a,c) *t*=490 ps (a) and (b,d) *t*=520 ps for *NP*=2. Fluid movement is illustrated in (c) and (d) for *t*=490 ps and *t*=520 ps, respectively.

Figure 8: Temperature profile at (a,c) *t=490* ps (a) and (b,d) *t=520* ps for *NP=5*. Fluid movement is illustrated in (c) and (d) for *t=490* ps and *t=520* ps, respectively.

## C. Total Intensity profile

The laser intensity of the Gaussian profile is given by

$$I_{Gaussian} = Ae^{-4\ln 2\left(\frac{t-3\tau_p}{\tau_p}\right)^2} e^{-2\left(\frac{x^2+y^2}{R_0^2}\right)} = P_1 e^{-4\ln 2\left(\frac{t-3\tau_p}{\tau_p}\right)^2} \qquad (1)$$

where $P_1 \equiv Ae^{-2\left(\frac{x^2+y^2}{R_0^2}\right)}$, and $A$ contains the pulse duration and fluence [1],

By contrast, the DLIP intensity is provided by the expression

$$I_{DLIP} = BI_0^{(i)} e^{-4\ln 2\left(\frac{t-3\tau_p}{\tau_p}\right)^2} e^{-2\left(\frac{x^2+y^2}{R_0^2}\right)} = P_2 e^{-4\ln 2\left(\frac{t-3\tau_p}{\tau_p}\right)^2} \qquad (2)$$

where $P_2 \equiv BI_0^{(i)} e^{-2\left(\frac{x^2+y^2}{R_0^2}\right)}$ (*i*=2 or 4 for a DLIP with two or four laser pulses, respectively, as given in Eq.1 in the main text) while $B$ contains the pulse duration and fluence [2],

Assuming that there is a temporal separation between the two pulses (equal to *Δτ*), there are two possibilities for the total intensity that depends on which pulse irradiates the material first

$$I_{total}(t,x,y,surface) = \left[P_1 e^{-4\ln 2\left(\frac{t-3\tau_p}{\tau_p}\right)^2} + P_2 e^{-4\ln 2\left(\frac{t-3\tau_p-\Delta\tau}{\tau_p}\right)^2}\right] \qquad (3)$$

if the Gaussian pulse *precedes* the DLIP pulse

and

$$I_{total}(t,x,y,surface) = \left[P_2 e^{-4\ln 2\left(\frac{t-3\tau_p}{\tau_p}\right)^2} + P_1 e^{-4\ln 2\left(\frac{t-3\tau_p-\Delta\tau}{\tau_p}\right)^2}\right] \qquad (4)$$

if the Gaussian pulse *follows* the DLIP pulse.

We can write Eqs.3,4 in a compact form in which $G_1 = 0, G_2 = 1$ (Eq.3) and $G_1 = 1, G_2 = 0$ (Eq.4).

$$I_{total}(t,x,y,surface) = \left[ P_1 e^{-4\ln 2\left(\frac{t-3\tau_p - G_1 \Delta\tau}{\tau_p}\right)^2} + P_2 e^{-4\ln 2\left(\frac{t-3\tau_p - G_2 \Delta\tau}{\tau_p}\right)^2} \right] \qquad (5)$$

## D. Navier-Stokes Equations

Below, the Navier-Stokes equations (for an incompressible fluid $\vec{\nabla} \cdot \vec{u} = 0$) are presented in a matrix form in 3D [3]. We know that

$$\rho_0 \left( \frac{\partial \vec{u}}{\partial t} + \vec{u} \cdot \vec{\nabla}\vec{u} \right) = \vec{\nabla} \cdot \left( -P\mathbf{1} + \mu(\vec{\nabla}\vec{u}) + \mu(\vec{\nabla}\vec{u})^T \right) \qquad (6)$$

where

$$(\vec{\nabla}\vec{u}) = \begin{pmatrix} \frac{\partial u}{\partial x} & \frac{\partial v}{\partial x} & \frac{\partial w}{\partial x} \\ \frac{\partial u}{\partial y} & \frac{\partial v}{\partial y} & \frac{\partial w}{\partial y} \\ \frac{\partial u}{\partial z} & \frac{\partial v}{\partial z} & \frac{\partial w}{\partial z} \end{pmatrix}, (\vec{\nabla}\vec{u})^T = \begin{pmatrix} \frac{\partial u}{\partial x} & \frac{\partial u}{\partial y} & \frac{\partial u}{\partial z} \\ \frac{\partial v}{\partial x} & \frac{\partial v}{\partial y} & \frac{\partial v}{\partial z} \\ \frac{\partial w}{\partial x} & \frac{\partial w}{\partial y} & \frac{\partial w}{\partial z} \end{pmatrix}, P\mathbf{1} = \begin{pmatrix} P & 0 & 0 \\ 0 & P & 0 \\ 0 & 0 & P \end{pmatrix}, \text{ and } \vec{u} = (u,v,w)$$

After the calculations, we obtain the following formulae in matrix form (by taking into account that the fluid is incompressible)

$$\begin{aligned} \rho_0 \left( \frac{\partial u}{\partial t} + u\frac{\partial u}{\partial x} + v\frac{\partial u}{\partial y} + w\frac{\partial u}{\partial z} \right) &= -\frac{\partial P}{\partial x} + \mu \left( \frac{\partial^2 u}{\partial x^2} + \frac{\partial^2 u}{\partial y^2} + \frac{\partial^2 u}{\partial z^2} \right) \\ \rho_0 \left( \frac{\partial v}{\partial t} + u\frac{\partial v}{\partial x} + v\frac{\partial v}{\partial y} + w\frac{\partial v}{\partial z} \right) &= -\frac{\partial P}{\partial y} + \mu \left( \frac{\partial^2 v}{\partial x^2} + \frac{\partial^2 v}{\partial y^2} + \frac{\partial^2 v}{\partial z^2} \right) \\ \rho_0 \left( \frac{\partial w}{\partial t} + u\frac{\partial w}{\partial x} + v\frac{\partial w}{\partial y} + w\frac{\partial w}{\partial z} \right) &= -\frac{\partial P}{\partial z} + \mu \left( \frac{\partial^2 w}{\partial x^2} + \frac{\partial^2 w}{\partial y^2} + \frac{\partial^2 w}{\partial z^2} \right) \end{aligned} \qquad (7)$$

Eq.7 represents the standard notation of NSE.

## E. Surface patterns

In the following figures, we illustrate enlarged contour plots of the surface patterns in which depth variation is shown (enlarged contour plots shown in last column of Figure 4 in the main text).

Figure 9: Surface contour plots for *NP*=50 for (a) *G+V*, (b) *V+G*, (c) *G+H*, (d) *H+G*, (e) *G+D*, (f) *G+D*

## F. Surface distribution of height

In the following figures, we illustrate surface patterns in which surface distribution of height is shown (in a 'blue to red' colorbar)

Figure 9: Spatial distribution of height for *NP*=50 for (a) *G+V*, (b) *V+G*, (c) *G+H*, (d) *H+G*, (e) *G+D*, (f) *G+D*

35